\documentclass[sigconf]{aamas}

\usepackage{balance} 

\setcopyright{ifaamas}
\acmConference[AAMAS '25]{Proc.\@ of the 24th International Conference
on Autonomous Agents and Multiagent Systems (AAMAS 2025)}{May 19 -- 23, 2025}
{Detroit, Michigan, USA}{A.~El~Fallah~Seghrouchni, Y.~Vorobeychik, S.~Das, A.~Nowe (eds.)}
\copyrightyear{2025}
\acmYear{2025}
\acmDOI{}
\acmPrice{}
\acmISBN{}

\acmSubmissionID{<<submission id>>}

\title[Dynamic Coalition Structure Detection]{Dynamic Coalition Structure Detection in Natural Language-based Interactions}

\author{Abhishek N. Kulkarni*}
\affiliation{
  \institution{University of Texas at Austin}\thanks{* Equal contribution.}
  \city{Austin}
  \country{USA}}
\email{abhishek.kulkarni@austin.utexas.edu}

\author{Andy Liu*}
\affiliation{
  \institution{Carnegie Mellon University}
  \city{Pittsburgh}
  \country{USA}}
\email{andyliu@cs.cmu.edu}

\author{Jean-Rapha\"el Gaglione}
\affiliation{
   \institution{University of Texas at Austin}
  \city{Austin}
  \country{USA}}
\email{jr.gaglione@utexas.edu}

\author{Daniel Fried}
\affiliation{
  \institution{Carnegie Mellon University}
  \city{Pittsburgh}
  \country{USA}}
\email{dfried@cs.cmu.edu}

\author{Ufuk Topcu}
\affiliation{
   \institution{University of Texas at Austin}
  \city{Austin}
  \country{USA}}
\email{utopcu@utexas.edu}

\usepackage{todonotes}
\usepackage{acronym}
\usepackage{amsmath}
\usepackage{amsthm}
\usepackage{mathtools}
\usepackage{graphicx}
\usepackage{cite}
\usepackage{bbm}
\usepackage[commandnameprefix=always]{changes}
\usepackage{booktabs} 
\usepackage{placeins}

\acrodef{mdp}[MDP]{Markov decision process}
\acrodef{pomdp}[POMDP]{Partially Observable Markov Decision Process}
\acrodef{momdp}[MOMDP]{Multi-objective MDP}
\acrodef{dfa}[DFA]{deterministic finite automaton}
\acrodef{tlmdp}[TLMDP]{terminating labeled Markov decision process}
\acrodef{lmdp}[LMDP]{labeled Markov decision process}
\acrodef{pdfa}[PDFA]{preference deterministic finite automaton}
\acrodef{pdra}[PDRA]{preference deterministic Rabin automaton}
\acrodef{cpltlf}[CPLTL$_f$]{Conditional Preference over LTL$_f$}
\acrodef{cpa}[CPA]{Conditional Preference Automaton}
\acrodef{ltl}[LTL]{linear temporal logic}
\acrodef{ltlf}[LTL$_f$]{linear temporal logic over finite traces}

\newcommand{\dist}{\mathcal{D}}

\newcommand{\supp}{\mathsf{Supp}}

\newcommand{\prob}{\mathsf{Pr}}
\newcommand{\indicator}{\mathbf{1}}

\newcommand{\weight}{\mathsf{wt}}
\newcommand{\agreements}{\mathsf{Agmt}}

\theoremstyle{definition}

\newtheorem{definition}{Definition}
\newtheorem{problem}{Problem}

\newcommand{\refDef}[1]{Def.~\ref{#1}}
\newcommand{\refEq}[1]{Eq.~(\ref{#1})}

\newcommand{\refFig}[1]{Fig.~(\ref{#1})}

\newcommand{\refSec}[1]{Sec.~\ref{#1}}

\begin{abstract}
	
In strategic multi-agent sequential interactions, detecting dynamic coalition structures is crucial for understanding how self-interested agents coordinate to influence outcomes. However, natural-language-based interactions introduce unique challenges to coalition detection due to ambiguity over intents and difficulty in modeling players' subjective perspectives. We propose a new method that leverages recent advancements in large language models and game theory to predict dynamic multilateral coalition formation in Diplomacy, a strategic multi-agent game where agents negotiate coalitions using natural language. The method consists of two stages. The first stage extracts the set of agreements discussed by two agents in their private dialogue, by combining a parsing-based filtering function with a fine-tuned language model trained to predict player intents. In the second stage, we define a new metric using the concept of subjective rationalizability from hypergame theory to evaluate the expected value of an agreement for each player. We then compute this metric for each agreement identified in the first stage by assessing the strategic value of the agreement for both players and taking into account the subjective belief of one player that the second player would honor the agreement. We demonstrate that our method effectively detects potential coalition structures in online Diplomacy gameplay by assigning high values to agreements likely to be honored and low values to those likely to be violated. The proposed method provides foundational insights into coalition formation in multi-agent environments with language-based negotiation and offers key directions for future research on the analysis of complex natural language-based interactions between agents.
\end{abstract}

\keywords{Coalition Structures, Game Theory, Multi-Agent Cooperation, Large Language Models}

\newcommand{\BibTeX}{\rm B\kern-.05em{\sc i\kern-.025em b}\kern-.08em\TeX}

\begin{document}


\pagestyle{fancy}
\fancyhead{}


\maketitle


\section{Introduction}
\label{sec:introduction}
The process of coalition formation in multi-agent systems involves agents forming coalitions to work together towards aligned objectives by coordinating their actions \citep{sandholm1999coalition,shehory1998methods}. 
This process has been studied extensively for both static and dynamic cases in game theory and logic.  
Past game-theoretic approaches have focused on studying which coalitions are likely to form based on various types of equilibria \citep{hajdukova2006coalition} and evaluating the value of a coalition to an agent. 
Meanwhile, logic-based approaches have focused on evaluating whether a given coalition can enforce a temporal property, regardless of how the agents not in the coalition behave \citep{pauly2002modal,alur2002alternating}.
However, neither of these approaches is suitable for studying coalition formation in natural language negotiation, where the ambiguity in language often leads players to interpret game states differently.
This phenomenon is commonly observed in real-world scenarios such as human-robot teaming \citep{chakraborti2016formal}, computer games \citep{mazrooei2013automating,rodriguez2022collusion}.

%
%

In this work, we study the problem of predicting dynamic multilateral coalition structures in sequential multi-agent interactions where players coordinate their actions using natural language. 
A coalition structure \citep{greenberg1994coalition} is a graph where nodes represent players and the edges represent agreements between players over coordinated actions.
While traditional approaches define coalition structures to be a partition of players, we model a coalition as a multi-graph allowing multiple agreements between two players, in addition to allowing a player to form bilateral agreements with multiple players simultaneously.
We model multi-agent interactions as reactive games, where the player can renegotiate their agreements in every round.  
Our aim is to predict these coalition structures from the perspective of an external observer, similar in spirit to an agency monitoring a computer network for anomalous behavior.


We use the board game Diplomacy \citep{calhamer1974invention} as a testbed for dynamic coalition structure detection over natural-language negotiations. Diplomacy is a seven-player board game that exemplifies key challenges in multi-agent systems research, combining semi-cooperative strategic dynamics with natural language-based negotiation. Players aim to control a majority of 34 supply centers on a map of Europe by coordinating the movement of units. While Diplomacy is a zero-sum game, players must negotiate strategic coalitions to support their own plans or counteract the moves of other players.


Diplomacy highlights three key challenges central to studying coalition formation in games with natural-language negotiations: \emph{decision-making under incomplete information}, \emph{reasoning based on mental models of opponents}, and \emph{multilateral negotiations}. 
Since the negotiations are pairwise and private, each player has incomplete information about the negotiations a second player has had with other players. 
As a result, players must anticipate others' actions without full knowledge of all agreements. This requires a player to construct a mental model of the other player's incentives to estimate the likelihood they would honor the agreement in addition to weighing their own incentives to honor the agreement. Lastly, since a player can simultaneously negotiate multiple agreements about the same unit with different players, a player must ultimately select a subset of agreements to honor based on the strategic advantage they offer to the player and the likelihood of them being honored by the player with whom the agreement is made.

\begin{figure*}
    \includegraphics[width=0.9\textwidth]{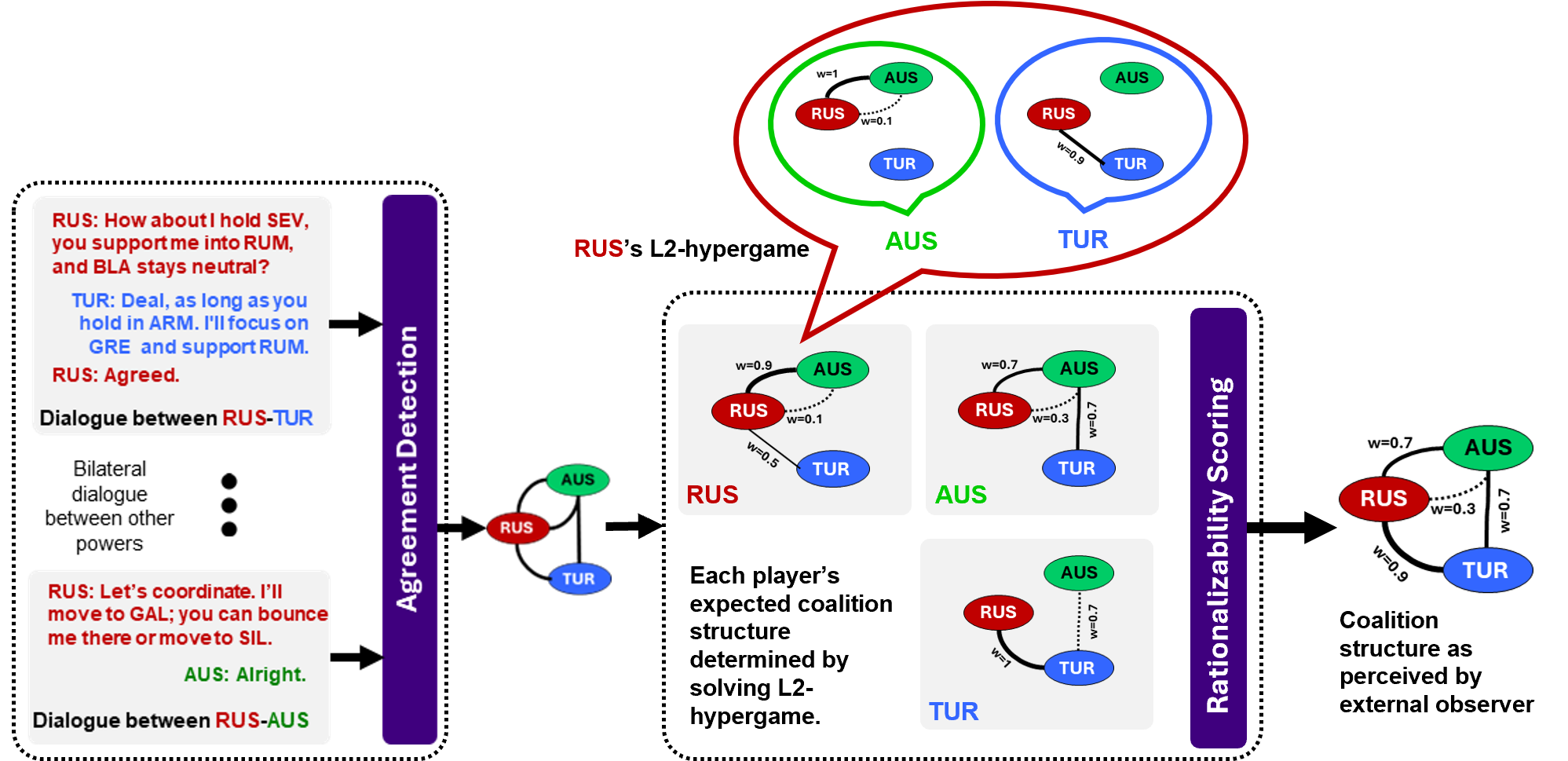}
    \caption{Proposed two-stage approach for learning coalition structures from natural language interactions in Diplomacy games. Stage 1 extracts agreements from pairwise dialogues to form an unweighted coalition structure. Stage 2 applies hypergame theory to assess the rationalizability of agreements for each player separately, which are then integrated into a weighted coalition structure representing the likelihood that an external observer believes agreements will be honored.}
    \label{fig:framework}
\end{figure*}

We define a novel method that addresses these challenges in dynamic coalition structure detection over natural language negotiations, as visualized in Figure~\ref{fig:framework}. Our approach consists of two stages: \textbf{agreement detection} and \textbf{strategic reasoning}.
To detect agreements negotiated via natural language in Diplomacy, we leverage large language models to parse dialogue, as well as  fine-tuned ``intent'' models from CICERO \citep{CICERO}, a Diplomacy-playing agent. By comparing the distribution over all moves involving parsed territories for a given unit before and after a phase of dialogue, we can learn whether a coalition was formed in the dialogue for that phase. Given a set of potential agreements identified, we then predict the set of honored agreements, which defines a coalition structure, using a deep reinforcement-learning based method. Due to the large action space and incomplete-information nature of Diplomacy, traditional enumerative game-theoretic approaches are intractable. To address these challenges, we extend the approach in \citep{bakhtin2021no} to compute the strategic value of an agreement for both players. We then sample from the intent model to determine the likelihood that both players will uphold the coalition, allowing us to measure the rationalizability of a coalition structure.

The three main contributions of this work are:
\begin{enumerate}
	\item \textbf{Approach.} We introduce a novel method that integrates large language models and game theory to predict dynamic multilateral coalition formation in multi-agent systems where agents negotiate coalitions using natural language.
	
	\item \textbf{Agreement detection.} We develop a procedure that combines pretrained language models with game dynamics to extract agreements from dialogues between agents, enabling the detection of coalition structures in real-time interactions.
	
	\item \textbf{Strategic reasoning.} We propose a new metric based on subjective rationalizability from hypergame theory, which evaluates the likelihood that agents will adhere to agreements by accounting for their subjective views of the game and strategic uncertainty.
\end{enumerate}

We validate our method on a dataset of online Diplomacy gameplay experiments.
We find that our hybrid agreement detection outperforms existing baselines and that our rationalizability metric effectively distinguishes between when players will honor coalition agreements and when they will not.
These findings highlight the value of integrating natural language techniques with game-theoretic analysis. 
They extend existing game-theoretic dynamic coalition prediction approaches to handle natural language negotiations, bridging toward more realistic real-world applications.


\subsection{Related Work} 

\textbf{Game theory.} 
The study of coalition formation in game theory focuses on identifying and characterizing stable coalitions by estimating the value of possible coalitions to an agent.
Solution concepts such as the core \citep{arnold2002dynamic}, the kernel \citep{shehory1996kernel}, the nucleolus \citep{montero2006noncooperative}, and the Shapley value \citep{aumann2003endogenous} have been introduced to analyze stability in transferable utility games, where side payments are allowed, and non-transferable utility games, where they are prohibited. 
However, these approaches do not account for sequential interactions where agents strategically join coalitions to achieve their goals.
The study of dynamic coalition formation in game theory has mainly focused on understanding the effects of externalities, where the formation of one coalition impacts the gains of other co-existing coalitions. \citep{rahwan2009coalition} presented a computational study of coalitional games with externalities, arguing that such externalities are common in real-world settings. The study of such externalities was extended by work including \citep{sklab2020coalition, skibski2016k, michalak2010logic}. While these approaches were able to better capture the impacts of externalities on other coalitions, they assume that complete information available to all agents, which is not applicable to games such as Diplomacy.

\textbf{Logic.} 
Strategic decision-making within coalitions has been studied within the logic community. Coalition Logic \citep{pauly2002modal} and Alternating-time Temporal Logic \citep{alur2002alternating} formalize reasoning about the existence of joint strategies for agents in coalitions to achieve their goals regardless of how the non-coalitional agents act. However, these logics only study static coalitions. 

There is an ongoing effort to extend coalition logic to handle dynamic settings. 
In \citep{umar2016coordinated}, authors introduce coordinated coalitions that represent a predefined sequence of coalitions for model checking.
\citep{guelev2023temporary} presents a complex framework that enriches Concurrent Game Models (CGM) by incorporating negotiations, where promises---represented as epistemic logic formulas---are embedded into states and existence of strategies that ensure goal satisfaction are verified through model checking. 
These methods requires full access to the game model, which is impractical for large-scale games like Diplomacy. 
Moreover, they model negotiations as deterministic statements, failing to capture the inherent ambiguity of natural language.

\textbf{Negotiation in Natural Language.} 
While significant past work has studied negotiation and coordination as a natural language task, the analysis of coalition formation in games involving natural language negotiations remains relatively understudied. \citep{lewis2017deal} collected a dataset of human negotiation dialogues in a semi-cooperative negotiation task, then trained natural language agents to perform the same task using the dataset. More recently, large language model-based agents have been used to achieve stronger performance on a range of social influence tasks \citep{chawla2023social}, including negotiation in a self-play environment over both zero-sum \citep{fu2023improving} and non-zero-sum \citep{liao2024efficacy} games. \citep{gandhi2023strategic} demonstrates that incorporating more explicit search and belief tracking into language models can improve their negotiation performance over a wide range of environments. \citep{moghimifar2024modelling} specifically seeks to model political coalition formation with language model-based agents, arguing that previous language model-based approaches to negotiation do not fully capture the full complexity and iterative nature of human negotiations. They contribute a multilingual dataset of European political party manifestos, as well as coalitions that they formed with other parties. While we similarly seek to model the multi-issue, iterative dynamics of natural-language coalition formation, we additionally analyze this problem from a game-theoretic perspective that accounts for agents' models of each other.

Diplomacy has also attracted attention from the natural language community as a testbed for the analysis of coordination dynamics in a strategic multi-agent environment. \citep{niculae-etal-2015-linguistic} studies the formation and termination of long-term alliances from a linguistic perspective, finding linguistic cues that presage acts of betrayal. \citep{peskov2020takes} models deception over long-term relationships in Diplomacy, finding that a model that uses both game dynamics and dialogue cues can predict player deception at a near-human level. \citep{wongkamjan2024more} analyzes games between CICERO and human Diplomacy players, noting that despite CICERO's strong strategic capabilities, it is still less persuasive compared to human players. Finally, \citep{mukobiwelfare} devises a novel positive-sum variant of Diplomacy, finding that language model-based agents are capable of attaining high joint welfare in this setting.

\section{Problem Formulation}
\label{sec:problem}
We use Diplomacy as a testbed to study dynamic multilateral coalition formation. 
Diplomacy is a deterministic game where players negotiate before concurrently submitting actions for their units, such as hold, move, or support.
The game state transitions based on these actions, and the negotiation phase repeats.
In this paper, we only consider the movement phases of the Diplomacy game. 
We note the high complexity of the game: each unit has an average of 26 valid orders, with up to 34 units on the board, making enumerative approaches intractable \citep{bakhtin2021no}.

\textbf{Game model.}  
Diplomacy can be modeled as a concurrent multiplayer game \citep{alur2002alternating} with rewards, $G = (N, S, A, T, s_0, R)$, where $N = \{P_i \mid i = 1, 2, \ldots n\}$ is a set of players, $S$ is a set of states, $A$ is a set of actions, $T: S \times A \rightarrow S$ is a deterministic transition function, $s_0 \in S$ is an initial state, and $R: S \times A \rightarrow \mathbb{R}$ is a reward function. 
A game play in $G$ is determined in two phases: 
Given a state $s \in S$, the players in $N$ \emph{privately} negotiate non-binding bilateral agreements with each other. 
\begin{definition}[Agreement]
	Given a state $s \in S$, an agreement between two players $P_i, P_j$ is a tuple $(u_1, u_2, a_1, a_2)$, where $u_1$ is a unit controlled by $P_i$, $u_2$ is a unit controlled $P_j$, and $a_1, a_2$ are legal actions for units $u_1, u_2$ in state $s$.
\end{definition}

The content of these negotiations and the agreements are only known to the players involved in the negotiation, unless one of these players explicitly shares this information with other players. 
At the conclusion of negotiation phase, all players choose an action, assigning an order to each unit controlled by them. 
Together, these actions determine the joint action $a = (a_1, a_2, \ldots, a_n)$, which in turn uniquely determines the next state $s' = T(s, a)$.
We denote by $d_t$ the set of all natural language messages exchanged between any pair of players in round $t$. 
Therefore, a game can be denoted as the sequence of state-dialogue-action pairs, $\rho = s_0 d_0 a_0 s_1 d_1 \ldots s_n$. 
A game in Diplomacy is of finite duration since a player will either win the game, or the game will be declared a draw.   
For a more detailed description, see \citep{sharp1978game}.

A policy for a player $P_i$ is a map $\pi_i: S \rightarrow \dist(A_i)$, where $ \dist(A_i)$ is a set of probability distributions over actions $A_i$ of player $P_i$.
A policy profile is a collection of policies of all players, $\pi = (\pi_1,...,\pi_n)$.

\textbf{Coalition structure.} A coalition is a set of honored agreements. 
The coalition structure, given a game state, is represented as an undirected multigraph with players as nodes and parallel edges indicating agreements between them. 
Since agreements are inferred from potentially ambiguous natural language dialogue, we assign weights to the edges.
Intuitively, these weights represent the likelihood of each agreement being honored.


\begin{definition}
	A coalition structure is a graph $C = (N, E, \agreements, \weight)$, where $N$ is the set of players, $\agreements$ is a set of agreements, $E \subseteq N \times N \times \agreements$ is the set of edges, and $\weight: E \rightarrow \mathbb{R}$ is a function that assigns a real-valued weight to each edge. 
\end{definition}

Note that in Diplomacy, the coalition structure is not static; we denote the coalition structure in round $t$ as $C_t$. This background motivates the problem of \textbf{coalition structure prediction}. 
\begin{problem}\label{eq:problem}
	Given a concurrent multiplayer game $G$, a round $t \geq 0$, and the play $\rho = s_0 d_0 a_0 \ldots s_t d_t$ until round $t$, predict the coalition structure $C_t$. 
\end{problem}

\section{Background: Hypergame Theory}
\label{sec:background}
Diplomacy is characterized by both \emph{incomplete information} and \emph{unawareness}. 
In Diplomacy, players make decisions without full knowledge, as they may be unaware of message exchanges between other players or the content of those messages. 
As a result, a player's rationality must be assessed based on their subjective view of the game, shaped by their knowledge (c.f. \citep{kulkarni2021synthesis,kulkarni2020deceptive}).


Hypergame is a game-theoretic model designed for games with incomplete information and unawareness  \citep{bennett1980hypergames,sasaki2012hypergames}.
In a hypergame, each player has a subjective view of their interaction, shaped by their knowledge of the game and others' perspectives.
This structure allows players to independently form subjective views and make decisions based on their own \emph{subjective game}, effectively capturing player unawareness within the model.

Formally, hypergames are defined inductively based on players' levels of perception. A level-0 (L0) hypergame represents a game with complete, symmetric information, where both players have the same perception of the game, identical to the true game. In a level-1 (L1) hypergame, at least one player misperceives the game, but neither player is aware of this discrepancy. Each player believes their perceptual game is the true game and plays accordingly, with these perceptual games being level-0 hypergames. In a level-2 (L2) hypergame, one player becomes aware of the misperception and can reason about the other player's perceptual game. This concept can be extended to higher hypergame levels; however, in this paper, we restrict ourselves to L2-hypergames, which are most directly relevant to Diplomacy.

\textbf{Subjective rationalizability} \citep{sasaki2014subjective} is a solution concept for hypergames that evaluates the rationality of players' actions based on their subjective views of the game, considering their knowledge and beliefs about other players' perspectives and actions.

\begin{definition}[Subjective Rationalizability] \label{def:sr}
    Let $H^2 = \langle H_1^1, H_2^1 \rangle$ denote a L2-hypergame, where $H_i^1 = (G_1^i, G_2^i)$ is player $P_i$'s L1-hypergame and $G_1^i$ is the subjective game of $P_1$ as perceived by $P_i$. 
    Then, a strategy $\pi_i^{\ast,2}$ is said to be subjectively rationalizable for player $P_2$ if and only if it satisfies the following condition for all $\pi_i$: 
    \[
	   u^2_i(\pi_i^{\ast,2},\pi^{\ast,2}_{j}, x) \ge u_i^2(\pi_i,\pi^{\ast,2}_{j}, x),
	\]
    %
	where $(i,j)\in \{(1,2), (2,1)\}$
	%
	%
    and $x$ is a distribution over $\Phi$ representing $P_2$'s hypothesis over some aspect of $P_1$'s game.
    In this case, the utility is calculated based on the expectation, that is, $u_i^2(\pi_i,\pi_j,x) =\sum_{\varphi\in \Phi} x(\varphi) u_i^2(\pi_i,\pi_j,\varphi)$.
	The strategy $\pi_1^{\ast,1}$ is subjectively rationalizable for $P_1$ if and only if it satisfies the following condition for all $ \pi_1$,
	\[
	u_1^1(\pi_1^{\ast,1}, \pi_2^{\ast,2}, \varphi_1 ) \ge
	u_1^1(\pi_1, \pi_2^{\ast,2}, \varphi_1 ),
	\]
	where $\pi_2^{\ast,2}$ is subjectively rationalizable for $P_2$.
\end{definition}

\refDef{def:sr} enables evaluating when a player's strategy is rational within their own subjective view of the game. 
For $P_2$, a strategy is subjectively rationalizable if, given its information about $P_1$'s game ($H_2^1$), $P_2$ cannot improve their utility by choosing a different strategy. 
Specifically, $P_2$'s utility from its chosen strategy, given the other player's strategy and their own beliefs (represented by a distribution $x$), must be at least as high as the utility from any other strategy they might choose. 
Subjective rationalizability is understood similarly for $P_1$.

\section{Coalition Structure Prediction Methodology}
\label{sec:methodology}
We introduce a two-stage approach as shown in \refFig{fig:framework} that integrates recent developments in LLMs with subjective rationalizability in hypergames to solve the problem of dynamic coalition structure prediction, as defined in Section~\ref{sec:problem}.

The first stage identify the set of candidate agreements $\agreements_t$ being discussed given a play $\rho_t$. 
We assume that all agreements are discussed in natural language and no side channels exist for forming agreements. 
Letting $C_t$ be the coalition structure at round $t$, the set $\agreements_t$ determines the set of edges of $C_t$.

The second stage assigns weights to the edges in $\agreements_t$, referred to as the \emph{rationalizability score}. 
For an agreement $\alpha \in \agreements_t$, this score represents the likelihood that an external observer, with access to the full state, action, and dialogue history, believes that $\alpha$ will be honored by both players. 
To compute the rationalizability score $\weight_i(\alpha)$, we estimate the likelihood of a player $P_i$ honoring the agreement using a L2-hypergame constructed by filtering the messages to include only those exchanged between the two players involved in $\alpha$.

Formally, the rationalizability score of an agreement $\alpha$ for a player $P_i$ is computed by evaluating the strategic value (utility) of $\alpha$ for $P_i$ in its hypergame $H_i^1$. Formally, it is given by
\begin{align*}
	\weight_i(\alpha) = V_i(\alpha) * V_{j}^i(\alpha),
\end{align*}
where $V_i(\alpha)$ is the game-theoretic value of agreement for $P_i$ and $V_{j}^i(\alpha)$ is $P_i$'s belief about the likelihood of $P_j$ honoring the agreement.
Based on \refDef{def:sr}, this weight reflects how subjectively rationalizable an agreement is for $P_i$, with higher weights indicating greater rationalizability.

Given the rationalizability scores $\weight_i(\alpha)$ and $\weight_j(\alpha)$ of $P_i$ and $P_j$, respectively, the rationalizability score of the agreement $\alpha$ for an external observer is given by 
\begin{align*}
	\weight(\alpha) = \weight_i(\alpha) * \weight_j(\alpha).
\end{align*}

Note that the proposed two stage approach separates the language-based reasoning from game-theoretic one.
The strategic values $V_i$ and $V_j^i$ are derived from game-theoretic solution concepts and do not rely on dialogue. 
Whereas, the set of agreements $\agreements_t$ is inferred directly from the dialogue.

In the remainder of this section, we first outline how agreements are identified from dialogue, followed by the computation of their rationalizability score.

\subsection{Agreement Detection}\label{sec:agreement}
To detect agreements in Diplomacy gameplay from game transcripts, we combine (1) a filtering stage where mentioned locations that a coalition can be formed over are extracted by a language model, and (2) an intent extraction stage where specialized Diplomacy models are used to extract player intents for classification. 

\begin{figure*}
    \includegraphics[width=0.8\textwidth]{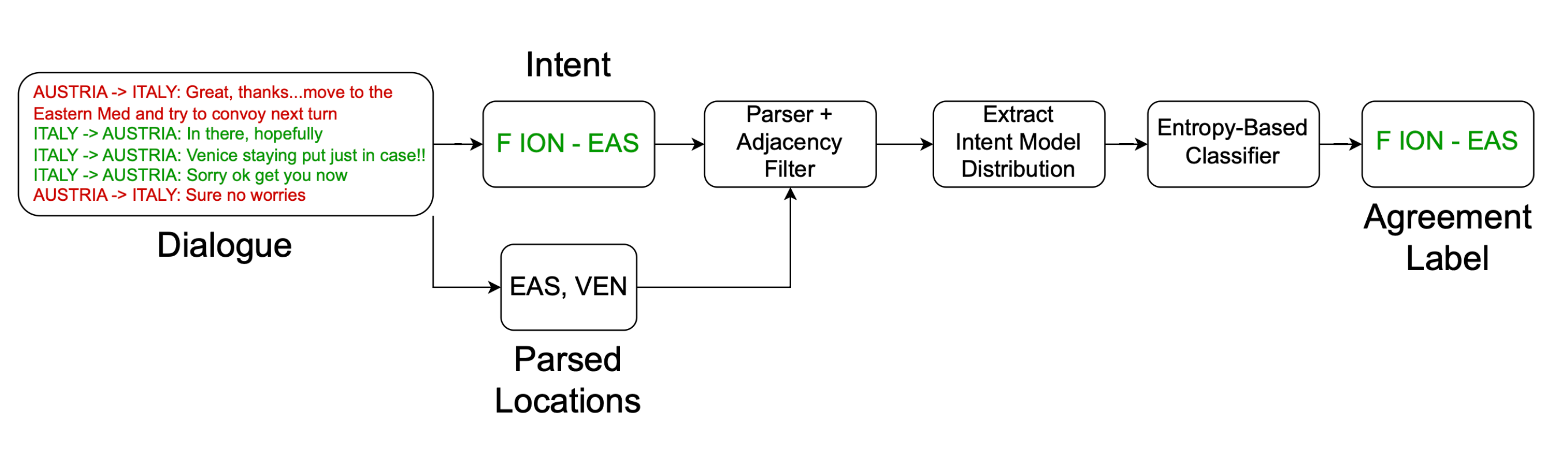}
    \caption{An overview of our agreement detection framework. In this case, we are analyzing whether Italy and Austria have come to an agreement over Italy's unit F ION, and determine that an agreement has been reached for Italy to move this unit to the Eastern Mediterranean Sea (EAS).}
    \label{fig:detection}
\end{figure*}

Figure~\ref{fig:detection} outlines our method for agreement detection. First, for each state-players tuple $(S, P_1, P_2)$, we first prompt GPT-4o\footnote{\url{https://platform.openai.com/docs/models/gpt-4o}}, a strong language model, with the dialogue between the two players at state $S$ and information about the Diplomacy board. We then use the model to extract all locations that were explicitly mentioned in negotiation between $P_1$ and $P_2$. In addition to information about whether two countries are sufficiently close to form a coalition, we will use this in a later filtering step. 

We then leverage the intent models used in CICERO \citep{CICERO}; these are 2.7-billion parameter language models that predict player actions from dialogue. Specifically, they are trained using behavioral cloning over a subset of ``truthful'' player dialogues collected from WebDiplomacy. Notably, this intent model only takes the conversation between $P_1$ and $P_2$ for a given phase, excluding any dialogue either player had with other players, to restrict the model to direct coordination between the two players.  By computing a distribution of move likelihoods over all possible moves for a unit before and after player dialogue, we can estimate whether a coalition was formed over the unit in question. We extract for each $(S, P_1, P_2, u \in u_1 \cup u_2)$ a most likely action $a^*$ for the unit in this state. We also compute the probability of $a^*$ before and after dialogue, as well as the entropy of the distribution of moves $P(a | S, P_1, P_2)$ before and after dialogue. After filtering out all units where a coalition is not possible, or where none of the territories involved in the move are mentioned in the dialogue, we then train a logistic regression classifier on these features to predict whether a coalition was formed over the unit in question.

This method allows us to leverage the advantages of both using a larger, general language model and a smaller, Diplomacy-specialized language model. While using the intent model allows us to capture more implicit coalition agreements that may not be identified with an explicit parser, it may also raise many false negatives due to noise in how the distribution changes as a result of unrelated dialogue. Adding a filtering step allows us to identify cases where the distribution shifts due to identifiable discussion of the provinces in question, as identified by the larger model. Indeed, in Section~\ref{sec:valid}, we show that this hybrid method outperforms methods that only rely on large language model annotation or learning from intent distributions.

\subsection{Strategic Value of Agreements}
%
%
%

Determining the strategic value $V_i(\alpha)$ of an agreement $\alpha = (u_1, u_2,$ $ a_1, a_2) \in \agreements$ for a player $P_i$ is a challenging task in large games like Diplomacy. 
It requires $P_i$ to determine the rational actions for all units controlled by $P_i$ as well as the other players conditioned on the unit $u_1$ being assigned action $a_1$ and the unit $u_2$ being assigned action $a_2$. 
Traditional game-theoretic approaches \citep{gibbons1992primer} enumerate all possible actions and evaluate them under a solution concept to determine the action that yields highest value from a given state. 
These approaches are inapplicable to games like Diplomacy due to the large size of players' action spaces.

Instead, we employ a deep reinforcement learning approach that first learns a probability distribution 
\begin{align} \label{eq:joint-action-distribution}
	\prob(a \mid s_0, \ldots, s_t, a_0, \ldots, a_{t-1},  \alpha)
\end{align} 
over joint actions of all players conditioned on $P_i$ and $P_j$ honoring a given agreement in addition to the state and action histories.
Intuitively, every joint action in the support of the distribution in \refEq{eq:joint-action-distribution} constitutes a Nash equilibrium in which $P_i$ and $P_j$ honor the agreement $\alpha$.


\textbf{Learning joint action distribution.} 
We leverage order sampling models trained as part of the CICERO agent \citep{CICERO}, which use Double oracle reinforcement learning for action exploration (DORA) \citep{bakhtin2021no} to learn the distribution in \refEq{eq:joint-action-distribution}. DORA simultaneously learns a state-value function and an joint action probability distribution using neural networks trained by bootstrapping on an approximate Nash equilibrium for the stage game each turn.

DORA is a Nash Q-Learning based approach to approximate Nash equilibrium in games with large state and action spaces.
It accommodates the large action spaces of Diplomacy by training a neural network $\pi(s; \theta_\pi)$ to predict joint action probability distribution with parameters $\theta_\pi$ that approximates the distribution of actions under the equilibrium policy at state $s$.
The candidate actions to explore are determined by sampling a large number of actions from $\pi(s; \theta_\pi)$ for each player and selecting actions with highest likelihood.
The Nash equilibrium is then estimated using regret minimization \citep{foster1999regret} in the matrix sub-game that includes only the sampled actions, assuming that the values of successor states are given by a learned network $\mathbf{V}(s; \theta_v)$, using the following update equation: 
\begin{align*}
	\mathbf{V}(s) \gets (1 - \beta) \mathbf{V}(s) + \beta(r(s) + \gamma \sigma(a) \mathbf{V}(T(s, a))).
\end{align*}


We refer interested readers to \citep{bakhtin2021no} for more details on the implementation. While we rely on CICERO-trained models for this work, versions of all of the specialized Diplomacy models used can be trained in novel game settings where human data is available \citep{bakhtin2021no}.

\textbf{Value of agreement.} 
The order sampling model and the value model provide a way to determine not only the joint action probabilities conditioned on an agreement, but also the value of the resulting state. 
Hence, we determine the value of an agreement by sampling from these distributions and computing the expected value of next state reached by the player by honoring $\alpha$,
\begin{align} \label{eq:value-of-agreement}
	V_i(\alpha) = \sum \prob(a \mid \vec{s}_t, \vec{a}_{t-1}, \alpha)  \mathbf{V}(s'),
\end{align}
where $s' = T(s, a)$ is the new state reached when joint action $a$ is performed in state $s$.

\subsection{Perceived Value of Agreement to Opponent}

While \refEq{eq:value-of-agreement} determines the value of an agreement for $P_i$, it does not allow $P_i$ to estimate its value for $P_j$ due to incomplete information about $P_j$'s negotiations with others. 
Instead, $P_i$ must infer $P_j$'s intent from their mutual dialogue and by estimating the value of various actions in the current state for $P_j$.


We interpret a players' intent as a probability distribution over the actions they assign to their units in the next round. 
To estimate the likelihood that $P_j$ will honor an agreement from $P_i$'s perspective, we approximate the intent distribution discussed in \refSec{sec:agreement}. 
This enables us to extract action probabilities from the dialogue to inform the agreement value computation.

Given the distribution in \refEq{eq:joint-action-distribution}, we compute the likelihood of $P_j$ respecting $\alpha$ using the following equation. 
We denote the support of a probability distribution $\mathbf{d}$ by $\supp(\mathbf{d})$. 
Given a joint action $a \in A$, let  $\indicator_j(a, \alpha) \mapsto \{\top, \bot\}$ denote whether the action $a$ assigns the same action with $P_j$'s unit as that assigned under $\alpha$. 
\begin{align*}
	V_j^i(\alpha) = \sum\limits_{a \in \supp(\mathbf{d})} \frac{\beta \prob(a \mid \vec{s}_t, \vec{d}_t, \vec{a}_{t-1})}{\beta \prob(a \mid \vec{s}_t, \vec{d}_t, \vec{a}_{t-1}) + (1 - \beta) \prob(a' \mid \vec{s}_t, \vec{d}_t, \vec{a}_{t-1})},
\end{align*}
where $\beta = \indicator_j(a, \alpha)$ and $a' \neq a$ is a valid action assigned to unit $u_2$ by $P_j$.

Intuitively, $V_j^i(\alpha)$ measures the relative value that $P_j$ achieves by selecting an action that honors $\alpha$ when compared with  selecting an action that does not honor $\alpha$, as players are more likely to honor agreements that are more strategically advantangeous.

\section{Experiments}

We evaluate the two stages of our proposed method separately. 
First, we outline the dataset employed for evaluation, and then  present the results for each of the two stages.

\subsection{Dataset}\label{sec:data}
We source previous Diplomacy games from WebDiplomacy\footnote{\url{https://webdiplomacy.net/}}, a multiplayer online implementation of Diplomacy. We consider a dataset of $140$ full-press games played over the standard Diplomacy map.  In order to calibrate our agreement detection classifier, we manually annotate five games from this dataset, randomly sampling from games with at least 250 total messages sent. This gives us a total of 16962 $(S, P_1, P_2, u)$ tuples over 1603 combinations of a state $S$ and players $P_1, P_2$. This dataset is highly imbalanced, with 444 (2.6\%) of all $(S, P_1, P_2, u)$ tuples having a coalition formed.

After we validate our usage of the agreement detection classifier, we then use it to label the remainder of the games with detected agreements. This resulting dataset consists of $415001$ total $(S, p_1, p_2, u)$ tuples. Of these tuples, $11008$ have agreements detected by our automatic method, $8344$ of which are upheld (i.e. the player played the agreed-upon move).

\subsection{Validating Agreement Detection Method}\label{sec:valid}

We use the manually-annotated data sample described in Section~\ref{sec:data} to test our agreement detection method, with an 80-20 train-test split. While this dataset is strongly imbalanced, we mitigate the impact of the dataset imbalance by only training on instances that pass our language model-based filter, which reduces our classifier training data to 1768 tuples, and by tuning a classification threshold to optimize F1-score on our training dataset. We benchmark three methods on this dataset: 
\begin{itemize}
    \item \textbf{GPT-4o}, which prompts a strong language model to directly identify units over which an agreement has been reached,
    \item \textbf{Classifier}, which trains a classifier on intent model distributions before and after dialogue over all $(S, P_1, P_2, u)$ tuples in the dataset, and
    \item \textbf{Hybrid}, our approach, which first filters using GPT-4o-parsed locations and player adjacency before training a classifier on the filtered data.
\end{itemize}

The results of our evaluation are in Table~\ref{tab:eval}. Our hybrid method outperforms both training a classifier on unfiltered intent data and prompting a strong language model on identifying units over which agreements have been reached. Extraction of coalition agreements from Diplomacy dialogue is a challenging task, due to the length of many dialogues in Diplomacy as well as the implicitness and fluctuating nature of negotiation over a multi-party dialogue. Fine-tuning more generally capable language models following the intent model formula in CICERO, in combination with more sophisticated parsers such as the one trained in \citep{wongkamjan2024more}, could yield even stronger performance improvements, which we leave to future work in this direction.

\begin{table}
    \centering
    \begin{tabular}{c|c|c|c}
    \toprule
    Method & F1 Score & Precision & Recall \\\midrule
    GPT-4o & 0.34 & 0.26 & 0.47 \\
    Classifier & 0.44 & 0.43 & 0.45 \\
    \textbf{Hybrid} & \textbf{0.55} & \textbf{0.63} & \textbf{0.48} \\\bottomrule
    \end{tabular}
    \caption{Classification metrics over the test dataset for our three methods. Hybrid methods outperform both purely language model-based and intent model distribution-based approaches at detecting whether an agreement has been reached over a specific unit.}
    \label{tab:eval}
\end{table}

\subsection{Evaluating Rationalizability Score}

The rationalizability score establishes a ranking of potential agreements for a unit within a specific game state. To evaluate the effectiveness of the score in predicting coalition structures, we analyze the rankings induced by the score on honored agreements in comparison to those of violated agreements.

We utilize both hand-labeled data and data labeled through the hybrid approach for the evaluation. 
We consider a total of $7434$ agreements labeled using the hybrid approach for evaluation.
For each agreement identified in the agreement detection stage, we generate a set of alternative agreements by sampling different orders for the units involved in the agreement.
The results of this evaluation for honored and violated agreements are presented in Table~\ref{tab:agreement}.
Given that the output of our model is a ranked list based on the rationalizability score, we employ two information retrieval metrics: mean reciprocal rank (MRR) \citep{craswell2009mean} and Brier score \citep{brier1950verification}.
The MRR is calculated using both the top-$1$ and top-$5$ ranked elements.

The ranking generated by the rationalizability score effectively differentiates between honored and violated agreements.
Our findings indicate that honored agreements typically receive lower ranks, while violated agreements tend to rank higher. 
This is observed through both the MRR and Brier scores.
When calculating the Brier score, we normalize the rationalizability scores, such that a score close to $1$ reflects that honored agreements usually have low ranks and violated agreements have higher ranks.
Notably, MRR scores that are close to $1$ in the top-$1$ case suggest that honored agreements are frequently assigned a rank of $0$, suggesting that this metric can very precisely recognize upheld coalitions.

We also compare our rationalizability score to a more conventional coalition formation prediction method, approximate Nash equilibrium (as estimated by the CICERO value model). We find that even when such approximate equilibrium-based methods are adapted for games with large state and action spaces, they remain inadequate for predicting coalition formation in such dynamic environments. Our R-Score yields a significantly higher MRR and a lower Brier score than the value model score in all cases.  This suggests that our rationalizability framework is significantly better at distinguishing between coalitions that are upheld and coalitions that are not upheld than Nash approximation-based predictions.

\begin{table}
\begin{centering}
\resizebox{0.5\textwidth}{!}{
\begin{tabular}{cc|ccc|ccc}
\toprule 
\multicolumn{2}{c}{} & \multicolumn{3}{c}{\textbf{Hand-Labelled}} & \multicolumn{3}{c}{\textbf{Hybrid}} \\
\multicolumn{1}{c}{Honored?} & \multicolumn{1}{c}{Metric} & \multicolumn{1}{c}{Value} & \multicolumn{1}{c}{R-Score@1} & \multicolumn{1}{c}{R-Score@5} & \multicolumn{1}{c}{Value} & \multicolumn{1}{c}{R-Score@1} & \multicolumn{1}{c}{R-Score@5} \\
\midrule
Yes & MRR ($\uparrow$) & 0.2842 & 0.9444 & \textbf{0.9722} & 0.3602 & 0.7416 & \textbf{0.8294} \\
No & MRR ($\downarrow$) & 0.2583 & \textbf{0.0} & 0.125 & 0.3682 & \textbf{0.2628} & 0.3354 \\
Yes & Brier ($\downarrow$) & 0.0802 & \textbf{0.0422} & \textbf{0.0422} & 0.0739 & \textbf{0.0311} & \textbf{0.0311} \\
No & Brier ($\uparrow$) & 0.7303 & \textbf{0.7494} & \textbf{0.7494} & 0.5706 & \textbf{0.6145} & \textbf{0.6145} \\
\bottomrule
\end{tabular}}
\end{centering}
\caption{Evaluation metrics for honored and violated agreements based on hand-labeled and hybrid datasets. The ranking induced by rationalizability score (RScore) on the set of agreements assigns lower ranks to honored agreements and higher ranks to violated agreements when compared to the ranking induced by Nash approximation-based predictions.}
\label{tab:agreement}
\end{table}

\section{Conclusion}

The detection of dynamic coalition structures is a key problem in understanding sequential interactions in strategic multi-agent environments. While many such environments use language, the study of coalition structure detection over natural language-based coordination is relatively understudied. This is compounded in settings like the board game Diplomacy, where players make decisions with incomplete information using dialogue-informed mental models of their opponents' future actions, and where relationships between players can shift drastically between turns as information is revealed.

Drawing from hypergame theory and the concept of subjective rationalizability, we propose a general method to dynamically predict coalition structures over sequential multi-agent interactions. In our method, we first extract detected agreements using the combination of a large language model-based parser and a specialized language model to predict player intents before and after the negotiation phase. We then compute the value of the agreement using a deep reinforcement-learning based value function, which we use in combination with player intents to compute the likelihood that each player will honor the agreement.

We validate the success of our method over sampled interactions between human Diplomacy players, using components of Meta's CICERO agent to compute player intents and action values. When compared to approximate Nash Equilibrium-based methods, our rationalizability score is significantly better at predicting the coalition structure at a given timestep. Our method can also generalize to other multi-agent, dialogue-based games, as long as sufficient human data exists upon which similar, game-specific models can be trained.

Extending coalition structure detection to natural language-based negotiation environments such as Diplomacy presents unique challenges in a setting where agents have incomplete information, negotiations are both multi-issue and multi-party, and where agents must reason over mental models of their opponents. However, for artificial agents to handle such complex environments properly, they must be capable of understanding the coalition dynamics of the environment at a given state. Our method and experiments serve as an important first step in this direction. We hope that future work will be able to extend our framework to new settings, including those with more complex negotiations and less existing domain-specific models, paving the way for agents that can reason over such information in deployment settings.

\FloatBarrier

\begin{acks}
This material is based upon work supported by the Defense Advanced Research Projects Agency (DARPA) under Agreement No. HR00112490410, the Army Research Lab under Agreement ARO W911NF-23-1-0317 and the Office of Naval Research under Agreement N00014-24-1-2097. We thank WebDiplomacy for supporting this research by providing access to online gameplay data.
\end{acks}





\bibliographystyle{ACM-Reference-Format}
\bibliography{main}


\begin{thebibliography}{42}


\ifx \showCODEN    \undefined \def \showCODEN     #1{\unskip}     \fi
\ifx \showDOI      \undefined \def \showDOI       #1{#1}\fi
\ifx \showISBNx    \undefined \def \showISBNx     #1{\unskip}     \fi
\ifx \showISBNxiii \undefined \def \showISBNxiii  #1{\unskip}     \fi
\ifx \showISSN     \undefined \def \showISSN      #1{\unskip}     \fi
\ifx \showLCCN     \undefined \def \showLCCN      #1{\unskip}     \fi
\ifx \shownote     \undefined \def \shownote      #1{#1}          \fi
\ifx \showarticletitle \undefined \def \showarticletitle #1{#1}   \fi
\ifx \showURL      \undefined \def \showURL       {\relax}        \fi
\providecommand\bibfield[2]{#2}
\providecommand\bibinfo[2]{#2}
\providecommand\natexlab[1]{#1}
\providecommand\showeprint[2][]{arXiv:#2}

\bibitem[\protect\citeauthoryear{Alur, Henzinger, and Kupferman}{Alur et~al\mbox{.}}{2002}]%
        {alur2002alternating}
\bibfield{author}{\bibinfo{person}{Rajeev Alur}, \bibinfo{person}{Thomas~A Henzinger}, {and} \bibinfo{person}{Orna Kupferman}.} \bibinfo{year}{2002}\natexlab{}.
\newblock \showarticletitle{Alternating-time temporal logic}.
\newblock \bibinfo{journal}{\emph{Journal of the ACM (JACM)}} \bibinfo{volume}{49}, \bibinfo{number}{5} (\bibinfo{year}{2002}), \bibinfo{pages}{672--713}.
\newblock


\bibitem[\protect\citeauthoryear{Arnold and Schwalbe}{Arnold and Schwalbe}{2002}]%
        {arnold2002dynamic}
\bibfield{author}{\bibinfo{person}{Tone Arnold} {and} \bibinfo{person}{Ulrich Schwalbe}.} \bibinfo{year}{2002}\natexlab{}.
\newblock \showarticletitle{Dynamic coalition formation and the core}.
\newblock \bibinfo{journal}{\emph{Journal of economic behavior \& organization}} \bibinfo{volume}{49}, \bibinfo{number}{3} (\bibinfo{year}{2002}), \bibinfo{pages}{363--380}.
\newblock


\bibitem[\protect\citeauthoryear{Aumann and Myerson}{Aumann and Myerson}{2003}]%
        {aumann2003endogenous}
\bibfield{author}{\bibinfo{person}{Robert~J Aumann} {and} \bibinfo{person}{Roger~B Myerson}.} \bibinfo{year}{2003}\natexlab{}.
\newblock \bibinfo{booktitle}{\emph{Endogenous formation of links between players and of coalitions: An application of the Shapley value}}.
\newblock \bibinfo{publisher}{Springer}.
\newblock


\bibitem[\protect\citeauthoryear{Bakhtin, Wu, Lerer, and Brown}{Bakhtin et~al\mbox{.}}{2021}]%
        {bakhtin2021no}
\bibfield{author}{\bibinfo{person}{Anton Bakhtin}, \bibinfo{person}{David Wu}, \bibinfo{person}{Adam Lerer}, {and} \bibinfo{person}{Noam Brown}.} \bibinfo{year}{2021}\natexlab{}.
\newblock \showarticletitle{No-press diplomacy from scratch}.
\newblock \bibinfo{journal}{\emph{Advances in Neural Information Processing Systems}}  \bibinfo{volume}{34} (\bibinfo{year}{2021}), \bibinfo{pages}{18063--18074}.
\newblock


\bibitem[\protect\citeauthoryear{Bennett}{Bennett}{1980}]%
        {bennett1980hypergames}
\bibfield{author}{\bibinfo{person}{Peter~G Bennett}.} \bibinfo{year}{1980}\natexlab{}.
\newblock \showarticletitle{Hypergames: developing a model of conflict}.
\newblock \bibinfo{journal}{\emph{Futures}} \bibinfo{volume}{12}, \bibinfo{number}{6} (\bibinfo{year}{1980}), \bibinfo{pages}{489--507}.
\newblock


\bibitem[\protect\citeauthoryear{Brier}{Brier}{1950}]%
        {brier1950verification}
\bibfield{author}{\bibinfo{person}{Glenn~W Brier}.} \bibinfo{year}{1950}\natexlab{}.
\newblock \showarticletitle{Verification of forecasts expressed in terms of probability}.
\newblock \bibinfo{journal}{\emph{Monthly weather review}} \bibinfo{volume}{78}, \bibinfo{number}{1} (\bibinfo{year}{1950}), \bibinfo{pages}{1--3}.
\newblock


\bibitem[\protect\citeauthoryear{Calhamer}{Calhamer}{1974}]%
        {calhamer1974invention}
\bibfield{author}{\bibinfo{person}{Allan Calhamer}.} \bibinfo{year}{1974}\natexlab{}.
\newblock \showarticletitle{The invention of diplomacy}.
\newblock \bibinfo{journal}{\emph{Games \& Puzzles}}  \bibinfo{volume}{21} (\bibinfo{year}{1974}).
\newblock


\bibitem[\protect\citeauthoryear{Chakraborti, Talamadupula, Zhang, and Kambhampati}{Chakraborti et~al\mbox{.}}{2016}]%
        {chakraborti2016formal}
\bibfield{author}{\bibinfo{person}{Tathagata Chakraborti}, \bibinfo{person}{Kartik Talamadupula}, \bibinfo{person}{Yu Zhang}, {and} \bibinfo{person}{Subbarao Kambhampati}.} \bibinfo{year}{2016}\natexlab{}.
\newblock \showarticletitle{A formal framework for studying interaction in human-robot societies}. In \bibinfo{booktitle}{\emph{Workshops at the Thirtieth AAAI Conference on Artificial Intelligence}}.
\newblock


\bibitem[\protect\citeauthoryear{Chawla, Shi, Zhang, Lucas, Yu, and Gratch}{Chawla et~al\mbox{.}}{2023}]%
        {chawla2023social}
\bibfield{author}{\bibinfo{person}{Kushal Chawla}, \bibinfo{person}{Weiyan Shi}, \bibinfo{person}{Jingwen Zhang}, \bibinfo{person}{Gale Lucas}, \bibinfo{person}{Zhou Yu}, {and} \bibinfo{person}{Jonathan Gratch}.} \bibinfo{year}{2023}\natexlab{}.
\newblock \showarticletitle{Social Influence Dialogue Systems: A Survey of Datasets and Models For Social Influence Tasks}. In \bibinfo{booktitle}{\emph{Proceedings of the 17th Conference of the European Chapter of the Association for Computational Linguistics}}. \bibinfo{pages}{750--766}.
\newblock


\bibitem[\protect\citeauthoryear{Craswell}{Craswell}{2009}]%
        {craswell2009mean}
\bibfield{author}{\bibinfo{person}{Nick Craswell}.} \bibinfo{year}{2009}\natexlab{}.
\newblock \showarticletitle{Mean reciprocal rank}.
\newblock \bibinfo{journal}{\emph{Encyclopedia of database systems}} (\bibinfo{year}{2009}), \bibinfo{pages}{1703--1703}.
\newblock


\bibitem[\protect\citeauthoryear{(FAIR)†, Bakhtin, Brown, Dinan, Farina, Flaherty, Fried, Goff, Gray, Hu, Jacob, Komeili, Konath, Kwon, Lerer, Lewis, Miller, Mitts, Renduchintala, Roller, Rowe, Shi, Spisak, Wei, Wu, Zhang, and Zijlstra}{(FAIR)† et~al\mbox{.}}{2022}]%
        {CICERO}
\bibfield{author}{\bibinfo{person}{Meta Fundamental AI Research Diplomacy~Team (FAIR)†}, \bibinfo{person}{Anton Bakhtin}, \bibinfo{person}{Noam Brown}, \bibinfo{person}{Emily Dinan}, \bibinfo{person}{Gabriele Farina}, \bibinfo{person}{Colin Flaherty}, \bibinfo{person}{Daniel Fried}, \bibinfo{person}{Andrew Goff}, \bibinfo{person}{Jonathan Gray}, \bibinfo{person}{Hengyuan Hu}, \bibinfo{person}{Athul~Paul Jacob}, \bibinfo{person}{Mojtaba Komeili}, \bibinfo{person}{Karthik Konath}, \bibinfo{person}{Minae Kwon}, \bibinfo{person}{Adam Lerer}, \bibinfo{person}{Mike Lewis}, \bibinfo{person}{Alexander~H. Miller}, \bibinfo{person}{Sasha Mitts}, \bibinfo{person}{Adithya Renduchintala}, \bibinfo{person}{Stephen Roller}, \bibinfo{person}{Dirk Rowe}, \bibinfo{person}{Weiyan Shi}, \bibinfo{person}{Joe Spisak}, \bibinfo{person}{Alexander Wei}, \bibinfo{person}{David Wu}, \bibinfo{person}{Hugh Zhang}, {and} \bibinfo{person}{Markus Zijlstra}.} \bibinfo{year}{2022}\natexlab{}.
\newblock \showarticletitle{Human-level play in the game of <i>Diplomacy</i> by combining language models with strategic reasoning}.
\newblock \bibinfo{journal}{\emph{Science}} \bibinfo{volume}{378}, \bibinfo{number}{6624} (\bibinfo{year}{2022}), \bibinfo{pages}{1067--1074}.
\newblock
\urldef\tempurl%
\url{https://doi.org/10.1126/science.ade9097}
\showDOI{\tempurl}
\showeprint{https://www.science.org/doi/pdf/10.1126/science.ade9097}


\bibitem[\protect\citeauthoryear{Foster and Vohra}{Foster and Vohra}{1999}]%
        {foster1999regret}
\bibfield{author}{\bibinfo{person}{Dean~P Foster} {and} \bibinfo{person}{Rakesh Vohra}.} \bibinfo{year}{1999}\natexlab{}.
\newblock \showarticletitle{Regret in the on-line decision problem}.
\newblock \bibinfo{journal}{\emph{Games and Economic Behavior}} \bibinfo{volume}{29}, \bibinfo{number}{1-2} (\bibinfo{year}{1999}), \bibinfo{pages}{7--35}.
\newblock


\bibitem[\protect\citeauthoryear{Fu, Peng, Khot, and Lapata}{Fu et~al\mbox{.}}{2023}]%
        {fu2023improving}
\bibfield{author}{\bibinfo{person}{Yao Fu}, \bibinfo{person}{Hao Peng}, \bibinfo{person}{Tushar Khot}, {and} \bibinfo{person}{Mirella Lapata}.} \bibinfo{year}{2023}\natexlab{}.
\newblock \showarticletitle{Improving language model negotiation with self-play and in-context learning from ai feedback}.
\newblock \bibinfo{journal}{\emph{arXiv preprint arXiv:2305.10142}} (\bibinfo{year}{2023}).
\newblock


\bibitem[\protect\citeauthoryear{Gandhi, Sadigh, and Goodman}{Gandhi et~al\mbox{.}}{[n.d.]}]%
        {gandhi2023strategic}
\bibfield{author}{\bibinfo{person}{Kanishk Gandhi}, \bibinfo{person}{Dorsa Sadigh}, {and} \bibinfo{person}{Noah Goodman}.} \bibinfo{year}{[n.d.]}\natexlab{}.
\newblock \showarticletitle{Strategic Reasoning with Language Models}. In \bibinfo{booktitle}{\emph{NeurIPS 2023 Foundation Models for Decision Making Workshop}}.
\newblock


\bibitem[\protect\citeauthoryear{Gibbons et~al\mbox{.}}{Gibbons et~al\mbox{.}}{1992}]%
        {gibbons1992primer}
\bibfield{author}{\bibinfo{person}{Robert Gibbons} {et~al\mbox{.}}} \bibinfo{year}{1992}\natexlab{}.
\newblock \showarticletitle{A primer in game theory}.
\newblock  (\bibinfo{year}{1992}).
\newblock


\bibitem[\protect\citeauthoryear{Greenberg}{Greenberg}{1994}]%
        {greenberg1994coalition}
\bibfield{author}{\bibinfo{person}{Joseph Greenberg}.} \bibinfo{year}{1994}\natexlab{}.
\newblock \showarticletitle{Coalition structures}.
\newblock \bibinfo{journal}{\emph{Handbook of game theory with economic applications}}  \bibinfo{volume}{2} (\bibinfo{year}{1994}), \bibinfo{pages}{1305--1337}.
\newblock


\bibitem[\protect\citeauthoryear{Guelev}{Guelev}{2023}]%
        {guelev2023temporary}
\bibfield{author}{\bibinfo{person}{Dimitar~P Guelev}.} \bibinfo{year}{2023}\natexlab{}.
\newblock \showarticletitle{Of Temporary Coalitions in Terms of Concurrent Game Models, Announcements, and Temporal Projection}. In \bibinfo{booktitle}{\emph{International Workshop on Logic, Rationality and Interaction}}. Springer, \bibinfo{pages}{126--134}.
\newblock


\bibitem[\protect\citeauthoryear{Hajdukov{\'a}}{Hajdukov{\'a}}{2006}]%
        {hajdukova2006coalition}
\bibfield{author}{\bibinfo{person}{Jana Hajdukov{\'a}}.} \bibinfo{year}{2006}\natexlab{}.
\newblock \showarticletitle{Coalition formation games: A survey}.
\newblock \bibinfo{journal}{\emph{International Game Theory Review}} \bibinfo{volume}{8}, \bibinfo{number}{04} (\bibinfo{year}{2006}), \bibinfo{pages}{613--641}.
\newblock


\bibitem[\protect\citeauthoryear{Kulkarni and Fu}{Kulkarni and Fu}{2021}]%
        {kulkarni2021synthesis}
\bibfield{author}{\bibinfo{person}{Abhishek~N Kulkarni} {and} \bibinfo{person}{Jie Fu}.} \bibinfo{year}{2021}\natexlab{}.
\newblock \showarticletitle{Synthesis of deceptive strategies in reachability games with action misperception}. In \bibinfo{booktitle}{\emph{Proceedings of the Twenty-Ninth International Conference on International Joint Conferences on Artificial Intelligence}}. \bibinfo{pages}{217--223}.
\newblock


\bibitem[\protect\citeauthoryear{Kulkarni, Luo, Leslie, Kamhoua, and Fu}{Kulkarni et~al\mbox{.}}{2020}]%
        {kulkarni2020deceptive}
\bibfield{author}{\bibinfo{person}{Abhishek~N Kulkarni}, \bibinfo{person}{Huan Luo}, \bibinfo{person}{Nandi~O Leslie}, \bibinfo{person}{Charles~A Kamhoua}, {and} \bibinfo{person}{Jie Fu}.} \bibinfo{year}{2020}\natexlab{}.
\newblock \showarticletitle{Deceptive labeling: hypergames on graphs for stealthy deception}.
\newblock \bibinfo{journal}{\emph{IEEE Control Systems Letters}} \bibinfo{volume}{5}, \bibinfo{number}{3} (\bibinfo{year}{2020}), \bibinfo{pages}{977--982}.
\newblock


\bibitem[\protect\citeauthoryear{Lewis, Yarats, Dauphin, Parikh, and Batra}{Lewis et~al\mbox{.}}{2017}]%
        {lewis2017deal}
\bibfield{author}{\bibinfo{person}{Mike Lewis}, \bibinfo{person}{Denis Yarats}, \bibinfo{person}{Yann Dauphin}, \bibinfo{person}{Devi Parikh}, {and} \bibinfo{person}{Dhruv Batra}.} \bibinfo{year}{2017}\natexlab{}.
\newblock \showarticletitle{Deal or No Deal? End-to-End Learning of Negotiation Dialogues}. In \bibinfo{booktitle}{\emph{Proceedings of the 2017 Conference on Empirical Methods in Natural Language Processing}}. \bibinfo{pages}{2443--2453}.
\newblock


\bibitem[\protect\citeauthoryear{Liao, Tomlin, and Klein}{Liao et~al\mbox{.}}{2024}]%
        {liao2024efficacy}
\bibfield{author}{\bibinfo{person}{Austen Liao}, \bibinfo{person}{Nicholas Tomlin}, {and} \bibinfo{person}{Dan Klein}.} \bibinfo{year}{2024}\natexlab{}.
\newblock \showarticletitle{Efficacy of Language Model Self-Play in Non-Zero-Sum Games}.
\newblock \bibinfo{journal}{\emph{arXiv preprint arXiv:2406.18872}} (\bibinfo{year}{2024}).
\newblock


\bibitem[\protect\citeauthoryear{Mazrooei, Archibald, and Bowling}{Mazrooei et~al\mbox{.}}{2013}]%
        {mazrooei2013automating}
\bibfield{author}{\bibinfo{person}{Parisa Mazrooei}, \bibinfo{person}{Christopher Archibald}, {and} \bibinfo{person}{Michael Bowling}.} \bibinfo{year}{2013}\natexlab{}.
\newblock \showarticletitle{Automating collusion detection in sequential games}. In \bibinfo{booktitle}{\emph{Proceedings of the AAAI Conference on Artificial Intelligence}}, Vol.~\bibinfo{volume}{27}. \bibinfo{pages}{675--682}.
\newblock


\bibitem[\protect\citeauthoryear{Michalak, Marciniak, Szamotulski, Rahwan, Wooldridge, McBurney, and Jennings}{Michalak et~al\mbox{.}}{2010}]%
        {michalak2010logic}
\bibfield{author}{\bibinfo{person}{Tomasz Michalak}, \bibinfo{person}{Dorota Marciniak}, \bibinfo{person}{Marcin Szamotulski}, \bibinfo{person}{Talal Rahwan}, \bibinfo{person}{Michael Wooldridge}, \bibinfo{person}{Peter McBurney}, {and} \bibinfo{person}{Nicholas Jennings}.} \bibinfo{year}{2010}\natexlab{}.
\newblock \showarticletitle{A logic-based representation for coalitional games with externalities}.
\newblock  (\bibinfo{year}{2010}).
\newblock


\bibitem[\protect\citeauthoryear{Moghimifar, Li, Thomson, and Haffari}{Moghimifar et~al\mbox{.}}{2024}]%
        {moghimifar2024modelling}
\bibfield{author}{\bibinfo{person}{Farhad Moghimifar}, \bibinfo{person}{Yuan-Fang Li}, \bibinfo{person}{Robert Thomson}, {and} \bibinfo{person}{Gholamreza Haffari}.} \bibinfo{year}{2024}\natexlab{}.
\newblock \showarticletitle{Modelling Political Coalition Negotiations Using LLM-based Agents}.
\newblock \bibinfo{journal}{\emph{arXiv preprint arXiv:2402.11712}} (\bibinfo{year}{2024}).
\newblock


\bibitem[\protect\citeauthoryear{Montero}{Montero}{2006}]%
        {montero2006noncooperative}
\bibfield{author}{\bibinfo{person}{Maria Montero}.} \bibinfo{year}{2006}\natexlab{}.
\newblock \showarticletitle{Noncooperative foundations of the nucleolus in majority games}.
\newblock \bibinfo{journal}{\emph{Games and Economic Behavior}} \bibinfo{volume}{54}, \bibinfo{number}{2} (\bibinfo{year}{2006}), \bibinfo{pages}{380--397}.
\newblock


\bibitem[\protect\citeauthoryear{Mukobi, Erlebach, Lauffer, Hammond, Chan, and Clifton}{Mukobi et~al\mbox{.}}{[n.d.]}]%
        {mukobiwelfare}
\bibfield{author}{\bibinfo{person}{Gabriel Mukobi}, \bibinfo{person}{Hannah Erlebach}, \bibinfo{person}{Niklas Lauffer}, \bibinfo{person}{Lewis Hammond}, \bibinfo{person}{Alan Chan}, {and} \bibinfo{person}{Jesse Clifton}.} \bibinfo{year}{[n.d.]}\natexlab{}.
\newblock \showarticletitle{Welfare Diplomacy: Benchmarking Language Model Cooperation}. In \bibinfo{booktitle}{\emph{Socially Responsible Language Modelling Research}}.
\newblock


\bibitem[\protect\citeauthoryear{Niculae, Kumar, Boyd-Graber, and Danescu-Niculescu-Mizil}{Niculae et~al\mbox{.}}{2015}]%
        {niculae-etal-2015-linguistic}
\bibfield{author}{\bibinfo{person}{Vlad Niculae}, \bibinfo{person}{Srijan Kumar}, \bibinfo{person}{Jordan Boyd-Graber}, {and} \bibinfo{person}{Cristian Danescu-Niculescu-Mizil}.} \bibinfo{year}{2015}\natexlab{}.
\newblock \showarticletitle{Linguistic Harbingers of Betrayal: A Case Study on an Online Strategy Game}. In \bibinfo{booktitle}{\emph{Proceedings of the 53rd Annual Meeting of the Association for Computational Linguistics and the 7th International Joint Conference on Natural Language Processing (Volume 1: Long Papers)}}, \bibfield{editor}{\bibinfo{person}{Chengqing Zong} {and} \bibinfo{person}{Michael Strube}} (Eds.). \bibinfo{publisher}{Association for Computational Linguistics}, \bibinfo{address}{Beijing, China}, \bibinfo{pages}{1650--1659}.
\newblock
\urldef\tempurl%
\url{https://doi.org/10.3115/v1/P15-1159}
\showDOI{\tempurl}


\bibitem[\protect\citeauthoryear{Pauly}{Pauly}{2002}]%
        {pauly2002modal}
\bibfield{author}{\bibinfo{person}{Marc Pauly}.} \bibinfo{year}{2002}\natexlab{}.
\newblock \showarticletitle{A modal logic for coalitional power in games}.
\newblock \bibinfo{journal}{\emph{Journal of logic and computation}} \bibinfo{volume}{12}, \bibinfo{number}{1} (\bibinfo{year}{2002}), \bibinfo{pages}{149--166}.
\newblock


\bibitem[\protect\citeauthoryear{Peskov and Cheng}{Peskov and Cheng}{2020}]%
        {peskov2020takes}
\bibfield{author}{\bibinfo{person}{Denis Peskov} {and} \bibinfo{person}{Benny Cheng}.} \bibinfo{year}{2020}\natexlab{}.
\newblock \showarticletitle{It takes two to lie: One to lie, and one to listen}. In \bibinfo{booktitle}{\emph{Proceedings of ACL}}.
\newblock


\bibitem[\protect\citeauthoryear{Rahwan, Michalak, Jennings, Wooldridge, and McBurney}{Rahwan et~al\mbox{.}}{2009}]%
        {rahwan2009coalition}
\bibfield{author}{\bibinfo{person}{Talal Rahwan}, \bibinfo{person}{Tomasz Michalak}, \bibinfo{person}{Nicholas Jennings}, \bibinfo{person}{Michael Wooldridge}, {and} \bibinfo{person}{Peter McBurney}.} \bibinfo{year}{2009}\natexlab{}.
\newblock \showarticletitle{Coalition structure generation in multi-agent systems with positive and negative externalities}.
\newblock  (\bibinfo{year}{2009}).
\newblock


\bibitem[\protect\citeauthoryear{Rodr{\'\i}guez, Rodr{\'\i}guez-Montequ{\'\i}n, Ballesteros-P{\'e}rez, Love, and Signor}{Rodr{\'\i}guez et~al\mbox{.}}{2022}]%
        {rodriguez2022collusion}
\bibfield{author}{\bibinfo{person}{Manuel J~Garc{\'\i}a Rodr{\'\i}guez}, \bibinfo{person}{Vicente Rodr{\'\i}guez-Montequ{\'\i}n}, \bibinfo{person}{Pablo Ballesteros-P{\'e}rez}, \bibinfo{person}{Peter~ED Love}, {and} \bibinfo{person}{Regis Signor}.} \bibinfo{year}{2022}\natexlab{}.
\newblock \showarticletitle{Collusion detection in public procurement auctions with machine learning algorithms}.
\newblock \bibinfo{journal}{\emph{Automation in Construction}}  \bibinfo{volume}{133} (\bibinfo{year}{2022}), \bibinfo{pages}{104047}.
\newblock


\bibitem[\protect\citeauthoryear{Sandholm, Larson, Andersson, Shehory, and Tohm{\'e}}{Sandholm et~al\mbox{.}}{1999}]%
        {sandholm1999coalition}
\bibfield{author}{\bibinfo{person}{Tuomas Sandholm}, \bibinfo{person}{Kate Larson}, \bibinfo{person}{Martin Andersson}, \bibinfo{person}{Onn Shehory}, {and} \bibinfo{person}{Fernando Tohm{\'e}}.} \bibinfo{year}{1999}\natexlab{}.
\newblock \showarticletitle{Coalition structure generation with worst case guarantees}.
\newblock \bibinfo{journal}{\emph{Artificial intelligence}} \bibinfo{volume}{111}, \bibinfo{number}{1-2} (\bibinfo{year}{1999}), \bibinfo{pages}{209--238}.
\newblock


\bibitem[\protect\citeauthoryear{Sasaki}{Sasaki}{2014}]%
        {sasaki2014subjective}
\bibfield{author}{\bibinfo{person}{Yasuo Sasaki}.} \bibinfo{year}{2014}\natexlab{}.
\newblock \showarticletitle{Subjective rationalizability in hypergames}.
\newblock  (\bibinfo{year}{2014}).
\newblock


\bibitem[\protect\citeauthoryear{Sasaki and Kijima}{Sasaki and Kijima}{2012}]%
        {sasaki2012hypergames}
\bibfield{author}{\bibinfo{person}{Yasuo Sasaki} {and} \bibinfo{person}{Kyoichi Kijima}.} \bibinfo{year}{2012}\natexlab{}.
\newblock \showarticletitle{Hypergames and Bayesian games: a theoretical comparison of the models of games with incomplete information}.
\newblock \bibinfo{journal}{\emph{Journal of Systems Science and Complexity}} \bibinfo{volume}{25}, \bibinfo{number}{4} (\bibinfo{year}{2012}), \bibinfo{pages}{720--735}.
\newblock


\bibitem[\protect\citeauthoryear{Sharp}{Sharp}{1978}]%
        {sharp1978game}
\bibfield{author}{\bibinfo{person}{Richard Sharp}.} \bibinfo{year}{1978}\natexlab{}.
\newblock \showarticletitle{The Game of Diplomacy}.
\newblock  (\bibinfo{year}{1978}).
\newblock


\bibitem[\protect\citeauthoryear{Shehory and Kraus}{Shehory and Kraus}{1996}]%
        {shehory1996kernel}
\bibfield{author}{\bibinfo{person}{Onn Shehory} {and} \bibinfo{person}{Sarit Kraus}.} \bibinfo{year}{1996}\natexlab{}.
\newblock \showarticletitle{A kernel-oriented model for coalition-formation in general environments: Implementation and results}. In \bibinfo{booktitle}{\emph{AAAI/IAAI, Vol. 1}}. \bibinfo{pages}{134--140}.
\newblock


\bibitem[\protect\citeauthoryear{Shehory and Kraus}{Shehory and Kraus}{1998}]%
        {shehory1998methods}
\bibfield{author}{\bibinfo{person}{Onn Shehory} {and} \bibinfo{person}{Sarit Kraus}.} \bibinfo{year}{1998}\natexlab{}.
\newblock \showarticletitle{Methods for task allocation via agent coalition formation}.
\newblock \bibinfo{journal}{\emph{Artificial intelligence}} \bibinfo{volume}{101}, \bibinfo{number}{1-2} (\bibinfo{year}{1998}), \bibinfo{pages}{165--200}.
\newblock


\bibitem[\protect\citeauthoryear{Skibski, Matejczyk, Michalak, Wooldridge, and Yokoo}{Skibski et~al\mbox{.}}{2016}]%
        {skibski2016k}
\bibfield{author}{\bibinfo{person}{Oskar Skibski}, \bibinfo{person}{Szymon Matejczyk}, \bibinfo{person}{Tomasz~P Michalak}, \bibinfo{person}{Michael~J Wooldridge}, {and} \bibinfo{person}{Makoto Yokoo}.} \bibinfo{year}{2016}\natexlab{}.
\newblock \showarticletitle{k-Coalitional Cooperative Games.}. In \bibinfo{booktitle}{\emph{AAMAS}}. \bibinfo{pages}{177--185}.
\newblock


\bibitem[\protect\citeauthoryear{Sklab, Aknine, Shehory, and Tari}{Sklab et~al\mbox{.}}{2020}]%
        {sklab2020coalition}
\bibfield{author}{\bibinfo{person}{Youcef Sklab}, \bibinfo{person}{Samir Aknine}, \bibinfo{person}{Onn Shehory}, {and} \bibinfo{person}{Abdelkamel Tari}.} \bibinfo{year}{2020}\natexlab{}.
\newblock \showarticletitle{Coalition formation with dynamically changing externalities}.
\newblock \bibinfo{journal}{\emph{Engineering Applications of Artificial Intelligence}}  \bibinfo{volume}{91} (\bibinfo{year}{2020}), \bibinfo{pages}{103577}.
\newblock


\bibitem[\protect\citeauthoryear{Umar and Mesbah}{Umar and Mesbah}{2016}]%
        {umar2016coordinated}
\bibfield{author}{\bibinfo{person}{Raza Umar} {and} \bibinfo{person}{Wessam Mesbah}.} \bibinfo{year}{2016}\natexlab{}.
\newblock \showarticletitle{Coordinated coalition formation in throughput-efficient cognitive radio networks}.
\newblock \bibinfo{journal}{\emph{Wireless Communications and Mobile Computing}} \bibinfo{volume}{16}, \bibinfo{number}{8} (\bibinfo{year}{2016}), \bibinfo{pages}{912--928}.
\newblock


\bibitem[\protect\citeauthoryear{Wongkamjan, Gu, Wang, Hermjakob, May, Stewart, Kummerfeld, Peskoff, and Boyd-Graber}{Wongkamjan et~al\mbox{.}}{2024}]%
        {wongkamjan2024more}
\bibfield{author}{\bibinfo{person}{Wichayaporn Wongkamjan}, \bibinfo{person}{Feng Gu}, \bibinfo{person}{Yanze Wang}, \bibinfo{person}{Ulf Hermjakob}, \bibinfo{person}{Jonathan May}, \bibinfo{person}{Brandon~M Stewart}, \bibinfo{person}{Jonathan~K Kummerfeld}, \bibinfo{person}{Denis Peskoff}, {and} \bibinfo{person}{Jordan~Lee Boyd-Graber}.} \bibinfo{year}{2024}\natexlab{}.
\newblock \showarticletitle{More Victories, Less Cooperation: Assessing Cicero's Diplomacy Play}.
\newblock \bibinfo{journal}{\emph{arXiv preprint arXiv:2406.04643}} (\bibinfo{year}{2024}).
\newblock


\end{thebibliography}


\end{document}